\documentstyle[12pt,epsfig]{article}
\begin{document}
\parskip=0pt
\parindent=125mm
HIP-1998-75/TH\par
TECHNION-PH-25\par
\today\par
\bigskip
\bigskip
\begin{center}
\parskip=20pt{\huge $b\to s s \bar d$ Decay in Two Higgs Doublet Models}
\bigskip
\bigskip
\bigskip
\bigskip
\bigskip

{\large Katri Huitu$^a$,  Cai-Dian L\"u$^{b,d}$\footnote{Alexander von Humboldt 
research fellow.}, 
Paul Singer$^c$ and Da-Xin Zhang$^{a,d}$ }

$^a$ 
Helsinki Institute of Physics, P.O.Box 9,
FIN-00014 University of Helsinki, Finland
\\
$^b$ II. Institut f\"ur Theoretische Physik, Universit\"at Hamburg,
22761 Hamburg, Germany
\\
$^c$ 
Department of Physics, Technion- Israel Institute of Technology,
Haifa 32000, Israel
\\
$^d$
The Abdus Salam International Centre for Theoretical Physics\\
P.O. Box 586, 34100 Trieste, Italy
\end{center}

\parindent=33pt
\bigskip
\begin{abstract}
 In the Standard Model, the box-diagram mediated decay $b\to s s \bar d$ is
predicted to occur with an extremely small branching ratio of below
$10^{-11}$,
thus providing a safe testing ground for exposing new physics.
We study this process in several Two Higgs Doublet Models and explore
their parameter space, 
indicating where this process becomes observable.
\end{abstract}


\newpage

\section{}
 In the study of possible virtual effects in B decays which are due to
physics beyond the Standard Model(SM), 
a crucial task is to subtract
reliably the SM contributions.
For the best studied rare $b$ decay process,
$b\to s+\gamma$,
many efforts have been made in the last decade to calculate
it as accurately as possible within SM.
However,
since the signals of new
physics in $b$ decays emerge unluckily with branching ratios at the level
of $10^{-7}$ or less \cite{fcnc-bey}, 
the commonly focused processes of
which $b\to s+\gamma$ is the prototype are very problematic for testing new
physics against the SM background.
Moreover,
other yet unobserved B decays,
like various $B\to\tau$ processes,
have also been shown to be rather
insensitive to a large class of new physics models \cite{nardi}.

In a recent letter \cite{hlsz},
we have emphasized an alternative approach to the
challenge of identifying virtual effects from new physics in $b$ decays,
by the consideration of processes which have negligible strength in the SM.
Such processes could thus serve as sensitive probes for new physics,
relatively free of SM ``pollution''.
In \cite{hlsz} we focused on the
$b\to ss\bar d$ transition,
which is a box-diagram induced process in the SM with
a very small branching ratio of below $10^{-11}$,
and we studied possible
effects from the minimal supersymmetric standard model (MSSM) and from a
supersymmetric model with R-parity violation.
We have shown that the 
existing limits on parameters of MSSM and of R-parity violating
interactions allow this process to occur well in excess of the SM rate,
thus providing an unusual opportunity for stricter limits or, hopefully,
for discovering new physics effects.
 In the present letter we undertake a study of the occurrence of the 
$ b\to ss\bar d$ transition in 
Two Higgs Doublet Models (THDMs)\cite{Hab,Gla,mod3},
which are
frequently considered as likely candidates for the extension of the SM.
We shall study two models in which the charged  Higgs exchange is
contributing to box diagrams,
which we denote as Model I \cite{Hab} and Model II \cite{Gla}, respectively,
as well as a third model allowing a tree level transition
mediated by neutral Higgs bosons \cite{mod3}.

We begin with some comments on the calculation of the W-box diagrams
in the SM.
Due to strong cancellations between the contributions from the top,
the charm and the up quarks in the loops, 
the leading order SM result
for the  $b\to ss\bar d$ decay rate is
\begin{eqnarray}
\Gamma=\frac{m_b^5}{48(2\pi)^3}
\left|\frac{G_F^2}{2\pi^2}m_W^2
V_{tb}V_{ts}^*\right|^2\left| V_{td}V_{ts}^* f\left(x\right)
+yV_{cd}V_{cs}^*
~g\left(x,y\right)\right|^2,\label{1}
\end{eqnarray}
where
\begin{eqnarray}
f(x)=\frac{1-11x+4x^2}{4x(1-x)^2}
-\frac{3}{2(1-x)^3}{\rm ln}x,
\end{eqnarray}
\begin{eqnarray}
g(x,y)= -\ln y +\frac{8x-4x^2-1}{4(1-x)^2}\ln x
+\frac{4x-1}{4(1-x)},
\end{eqnarray}
with $x=m_W^2/m_t^2$, $y=m_c^2/m_W^2$.
The first term in (\ref{1}) is suppressed by the small
Cabibbo-Kobayashi-Maskawa (CKM)
matrix elements $V_{td}V_{ts}^*$, 
while the second term is suppressed by $y=m_c^2/m_W^2$
and contributes about a half of the CKM suppressed term.
In this second contribution,
we have neglected a kinematics dependent term when
we perform the integral over the loop momentum.
This amounts to neglecting a small 
$(m_c^2/m_W^2)\ln (f(p)/m_c)$ contribution
wIth $f(p)$ a function depending on the external momenta $p$.
We have checked the effect of the neglected dependence numerically
and found that it never exceeds $10\%$ of the $m_c^2/m_W^2$ term
in the whole kinematic region\footnote{
We note that this dependence does not appear in
the calculation of the $K\bar K$ mixing, 
where the external momentum squares are of the order of $m_s^2$
and can be safely neglected compared to $m_c^2$,
nor in the case of $B\bar B$ mixing,
where even the contributions of
the order of $m_b^2/m_W^2$ are sub-dominant.}.
Since the involved CKM matrix elements are not well bounded and the
relative phase of the two terms is not fixed, we can only determine a range
for the branching ratio of $b\rightarrow ss\bar d$, which turns out to be 
always below $10^{-11}$ in the SM.
Note that QCD corrections may not change the value
by orders of magnitudes,
if compared with the analogous processes $B^0\bar B^0$ and
$K^0 \bar K^0$ mixing \cite{box}.
All these features combine to single out this process as a very sensitive
one to new physics.

The THDMs are the minimal extensions of the SM. 
With one more Higgs doublet,
one has to suppress the tree level flavor changing neutral
current (FCNC) interactions due to neutral Higgs bosons
to be consistent with the data.
The Model I allows only one Higgs
doublet to couple to both the up- and the down-type  quarks \cite{Hab}.
In the Model II one Higgs doublet is coupled only to up-type
quarks while the other doublet is coupled only to down-type quarks \cite{Gla},
and thus the Higgs content in the Model II is the same as in the MSSM.
In both the Model I and Model II,
discrete symmetries are enforced to forbid the more general couplings
and thus the tree level FCNC interactions are absent.
In another model, the Model III \cite{mod3},
the tree level FCNC is assumed to be small enough to 
lie within the experimental bounds.

\section{}

The relevant Lagrangian in the Model I and II is
\begin{eqnarray}
 {\cal L}&= &
\frac{1}{\sqrt{2}} \frac{g_2}{M_W}
\left[\cot \beta ~
{\bar u}_{i,R} (M_U V)_{ij} d_{j,L}-\xi{\bar u}_{i,L} (M_D V)_{ij} d_{j,R}
\right]H^+ ~+~ h.c.,\label{2hdl}
\end{eqnarray}
where $V$ represents the $3 \times 3$ unitary CKM matrix,
$M_U$ and $M_D$ denote the diagonalized quark mass matrices, 
the subscripts
$L$ and $R$ denote left-handed and right-handed quarks,
and $i,j=1,2,3$ are the generation indices.
For model I, $\xi=\cot \beta $; while for Model II, $\xi=-\tan \beta $.

\begin{figure}
    \epsfig{file=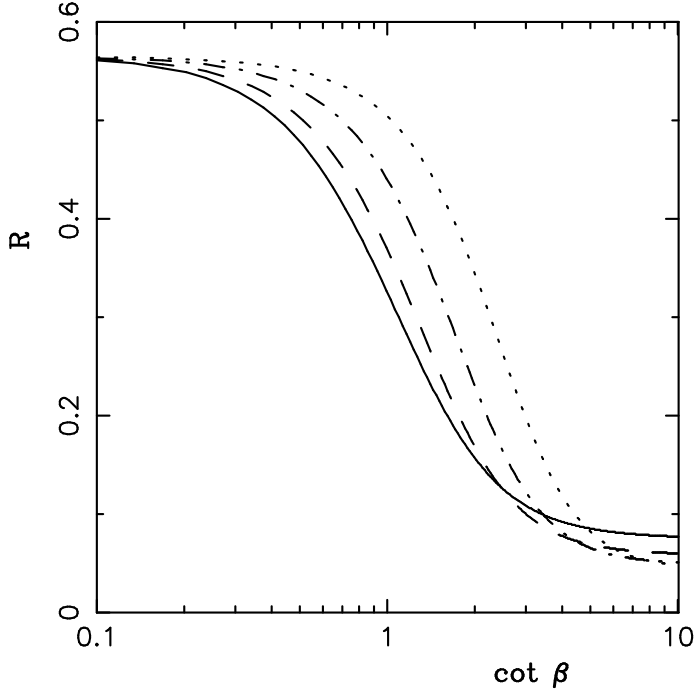,bbllx=2cm,bblly=7.5cm,bburx=21cm,bbury=19cm,%
width=14cm,angle=0}
    \caption{The ratio $R$
of the amplitude induced by the $m_c^2/m_W^2$ term
over that by the CKM suppressed one.
 The solid line, dashed line, dash-dotted line and dotted line
correspond to $m_{H^+}=100, 200, 400, 800$ GeV, respectively.}
    \label{fig0}
\end{figure}

In the charged Higgs mediated box diagrams, 
the couplings proportional to the up quark masses 
dominate the contribution to
$b\to s s\bar d $ decay. 
In both models we have
\begin{eqnarray}
\Gamma=\frac{m_b^5}{48(2\pi)^3}
\left|\frac{G_F^2}{2\pi^2}m_W^2 
V_{tb}V_{ts}^*\right|^2\left|V_{td}V_{ts}^*\left [f(x)+
A\left(x,z\right)\right]
+yV_{cd}V_{cs}^*\left [ g(x,y) +
B\left(x,z\right)\right ]\right|^2,\label{2hd}
\end{eqnarray}
where
\begin{eqnarray}
A(x,z)&=&\frac{\cot ^2 \beta}{2}
\left( \frac{1-4x}{x(1-z)(1-x)}+\frac{3x}{(1-x)^2(z-x)}\ln x 
+\frac{z^2-4xz}{x(1-z)^2(z-x)} \ln z \right)\nonumber\\
&+& 
\frac{\cot ^4 \beta }{4x} \left(\frac{1+z}{(1-z)^2}
+\frac{2z}{(1-z)^3}{\rm ln}z\right) ,
\end{eqnarray}
\begin{eqnarray}
B(x,z)&=&\cot ^2\beta \left(
\frac{4x-z}{2(z-x)(1-z)}\ln z 
-\frac{3x}{2(1-x)(z-x)}\ln x \right)\nonumber\\
&&-\frac{1}{4}\cot ^4 \beta \left(\frac{1}{1-z}
+\frac{1}{(1-z)^2} \ln z \right),
\end{eqnarray}
with $ z=m_{H^+}^2/m_t^2$. 
The functions $A(x,z)$ and $B(x,z)$ denote the new
contributions from the charged Higgs.
Because we have dropped in (\ref{2hd}) the terms
which are proportional 
to a factor less than $m_b m_s/M_W^2$
at the amplitude level,
the limit $\cot \beta\to 0$ corresponds to the SM case.
As in the SM,
the result depends on the relative CKM phase between the
terms with the $m_c^2/m_W^2$ and the CKM suppressed contributions.
In Fig.~1 we plot $R$ as the function of $\cot \beta$,
where
\begin{equation}
R=\frac{
|yV_{cd}V_{cs}^*\left [ g(x,y) +
B\left(x,z\right)\right ]|}{|V_{td}V_{ts}^*\left [f(x)+
A\left(x,z\right)\right]|}
\end{equation}
is the ratio between the amplitudes induced by the 
$m_c^2/m_W^2$ and by the CKM suppressed terms. 
It is clear that the SM is the limit with the maximal importance of
the $m_c^2/m_W^2$ contribution, 
while in the limit of large $\cot\beta$ this contribution
is negligible.

\begin{figure}
    \epsfig{file=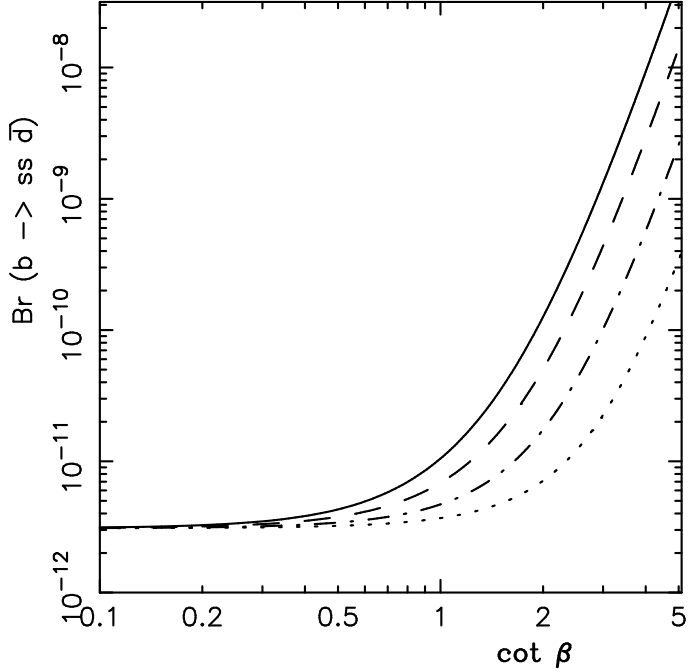,bbllx=2cm,bblly=7.5cm,bburx=21cm,bbury=19cm,%
width=14cm,angle=0}
    \caption{The branching ratio of $b\to ss\bar d$ as a function of 
$\cot \beta $. The solid line, dashed line, dash-dotted line and dotted line
correspond to $m_{H^+}=100, 200, 400, 800$ GeV, respectively.
Regions above the lines are excluded.}
    \label{fig}
\end{figure}

In the numerical calculation, 
we use $m_b=4.5$ GeV, $m_c=1.5$ GeV, $|V_{ts}|=0.04$,
$|V_{td}|=0.01$ and $|V_{cd}|=0.22$. 
We set the relative phase between the two
terms in (\ref{2hd}) as zero
which corresponds to a maximal constructive interference.
In Fig. 2, we show the numerical results for branching ratio of 
$b\to s s\bar d $ decay as a function of $\cot \beta$ for different 
charged Higgs masses. 
We have taken $m_{H^+}$ as $100$--$800$ GeV while
$\cot \beta$ ranging from 0.1 to 5.
Some semi-quantitative considerations
suggest a small $\cot\beta$ ($<2$),\footnote{We
refer the reader to \cite{soni} for more detailed discussions.}
in  which region the decay branching ratio for $b\to ss\bar d$ decay
is unobservable.
However, 
phenomenologically
a $\cot\beta$ as large as 5
is still consistent with the low energy data \cite{bhp}
and the direct search of the Higgs boson in top decays.
In this parameter space of a large $\cot\beta$
the charged Higgs box diagrams are sizable.
We note that in the MSSM $\cot\beta < 1$ is preferred
in model building,
corresponding to an unobservable contribution 
from the charged Higgs box diagrams in this model \cite{hlsz}.

As can be seen in Fig.~\ref{fig}, the branching ratio in the 
Model I and II can be of the order $10^{-8}$ at large $\cot \beta$ region. 
The searching of the decay $b\to s s \bar d$ and its hadronic channels 
$B^\pm\to K^\pm K^\pm X$ in B experiments 
will further constrain the parameters in the THDMs.

\section{}

In the THDM III,
there exists the Yukawa interaction \cite{mod3}
\begin{equation}
{\cal L}=\xi_{ij}{\bar Q}_{i,L}\phi_2 D_{j,R}~
+~({\rm the~up~quark~sector})~+~h.c..
\label{lag3}
\end{equation}
Here, 
the scalar Higgs doublet $\phi_2$ 
mediates the FCNC transitions $d_{i}\leftrightarrow d_{j}$ at the tree level,
if the coupling $\xi_{ij}$ is nonzero.
As discussed in the literature, FCNC effects induced by 
the loop diagrams are always negligible
compared to the tree level ones in the Model III \cite{ars}, 
hence the box diagrams with charged Higgs can be safely dropped
in the present decay.
We will neglect the unimportant QCD correction
since no GIM-like cancellation happens in the process.

The couplings in (\ref{lag3}) are constrained strongly from
the neutral meson mixing
by the requirement that the FCNC contribution to the mixing from
the interactions in (\ref{lag3})
does not exceed its experimental value.
In the $K \bar K$ and $B_s \bar B_s$ mixing, 
the dominant contributions
are proportional to $\xi_{sd}^2$ and $\xi_{sb}^2$, 
respectively \cite{ars}
\begin{equation}
\Delta M_F= 2 \xi_{sq}^2 \left( \frac{M_S^F}{m_h^2} +\frac{M_P^F}{m_A^2}
\right),
\end{equation}
with $F=K,B_S$, and $q=d,b$, respectively, and
\begin{eqnarray}
M_S^F &=& \frac{1}{6} \left( f_F^2 M_F 
+\frac{f_F^2M_F^3}{(m_s+m_q)^2}\right),
\nonumber \\
M_P^F &=& \frac{1}{6} \left( f_F^2 M_F 
+\frac{11f_F^2 M_F^3}{(m_s+m_q)^2}\right).
\end{eqnarray}
Here $m_h$ and $m_A$ are the masses of the neutral scalar
and pseudoscalar Higgs bosons.
Note that we have taken $\xi_{ij}=\xi_{ji}$
while it is not difficult to generalize to the 
case without this assumption.
In numerical calculations, 
we use $f_K= 160$ MeV, $f_{B_s}= 200$ MeV, 
$\Delta M_K= 3.491 \times 10^{-15}$ GeV \cite{pdg}.
Taking $m_A=m_h\equiv m_H$,
we obtain the bound
\begin{eqnarray}
\frac{\xi_{sd}}{m_H}< 8.3 \times 10^{-8} {\rm GeV}^{-1}
\end{eqnarray}
from $\Delta M_K$.
The experimental lower limit from $B_s-\bar B_s$ mixing
\begin{eqnarray}
\Delta M_{B_s} > 5.2 \times 10^{-12}{\rm GeV} ~\cite{aleph},
\label{bsbsbar}
\end{eqnarray}
gives no constraint on $\xi_{sb}/m_H$,
since the contribution in the SM itself exceeds this number.
Furthermore,
free from any assumption made for the FCNC Higgs 
couplings in the lepton sector,
the bounds from $B_s\to l_il_j$ and $B\to Kl_il_j$
($l_i,l_j=e,\mu ~{\rm or} ~\tau$)
do not exclude even $\xi_{sb}/m_H$ as large as
$10^{-1}$ \cite{sher}.

\begin{figure}
    \epsfig{file=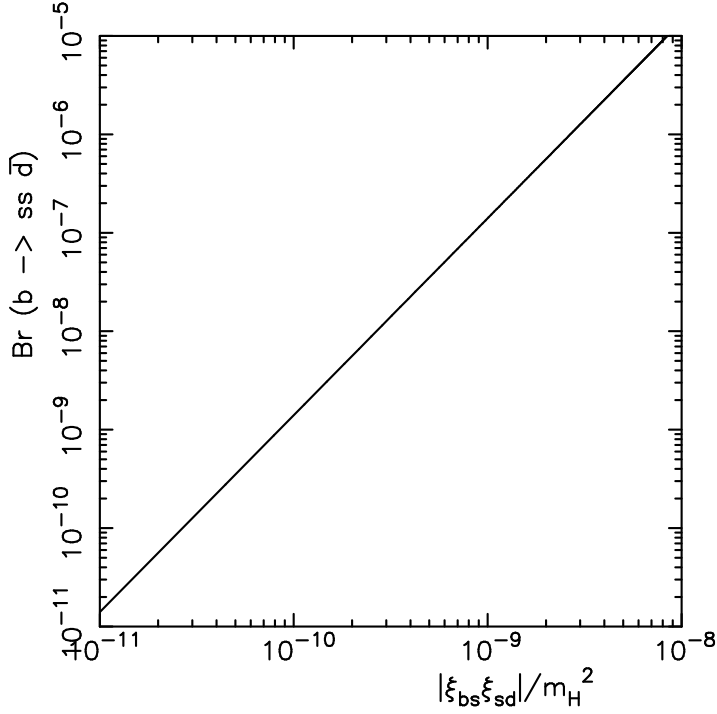,bbllx=2cm,bblly=7.5cm,bburx=21cm,bbury=19cm,%
width=14cm,angle=0}
    \caption{The branching ratio of $b\to ss\bar d$ in THDM III 
as a function of 
$|\xi_{bs}\xi_{sd}|/m_H^2$.}
    \label{fig3}
\end{figure}

In the presence of the interactions (\ref{lag3}),
$b\to ss\bar d$ can be induced by a tree diagram
exchanging the neutral Higgs bosons $h$ (scalar) and $A$
(pseudo-scalar), with the amplitude
\begin{equation}
{\cal A} = \frac{i}{2}~ \xi_{sb} \xi_{sd} \left( \frac{1}{m_h^2} 
(\bar s b)(\bar s d) - \frac{1}{m_A^2} (\bar s \gamma_5 b)(\bar s\gamma_5 d)
 \right) .
\end{equation}
The decay rate is thus
\begin{equation}
\Gamma= \frac{m_b^5}{3072(2\pi)^3} |\xi_{sb} \xi_{sd}|^2
\left\{ 11\left(\frac{1}{m_h^4}+\frac{1}{m_A^4}\right) +\frac{2}{m_h^2m_A^2}
 \right\}.\label{gamma3}
\end{equation}
The numerical results are shown in Fig.\ref{fig3} as a function of 
$|\xi_{sb} \xi_{sd}|/m_H^2$.

{}From Fig. 3,
it can be seen that
the decay $b\to ss\bar d$ is observable in the Model III, if
$|\xi_{sb} \xi_{sd}|/m_H^2>10^{-10}$ GeV$^{-2}$,
which corresponds to a branching ratio around $10^{-9}$.
This requires at least $|\xi_{sb}/m_H|>10^{-3}$ GeV$^{-1}$.
Corresponding to this number,
the neutral Higgs contribution 
to $\Delta M_{B_s}$ is $10^6$ times larger than its
present lower limit.
To our knowledge,
such a large  $\Delta M_{B_s}$ is difficult to exclude.

\section{}

In summary,
we have presented a study of the $b\to ss\bar d$ process in several THDMs.
Firstly,
we confirmed that the charged Higgs box contribution in
MSSM is indeed negligible \cite{hlsz},
while on the other hand in Models I \cite{Hab} and II \cite{Gla} 
this contribution can induce observable effects at the $10^{-8}$ level
for the branching ratio. 
A large $B_s \bar B_s$ mixing, 
which is at least $10^6$
times larger than its present lower limit, 
is required in Model 3 \cite{mod3} for
this process to be observable.
Combining the above results with those of Ref \cite{hlsz},
we reemphasize
that the search for the processes we recommended \cite{hlsz},
like $B^-\to K^-K^-\pi^+$,
will serve immediately to get better limits on the parameters of
R-parity violating theories; at the later stage, 
when lower experimental
limits are obtainable,
it would constitute a direct check of the various
Non-Standard Model theories investigated here and in Ref \cite{hlsz}.

The work of KH and DXZ is partially
supported by the Academy of Finland (no. 37599).
The work of PS is partially supported by the Fund for
Promotion of Research at the Technion.

\end{document}